# CORRECTION
# A CORRELATED TOPIC MODEL OF SCIENCE

*Ann. Appl. Statist.* **1** (2007) 17–35

By David M. Blei and John D. Lafferty

*Princeton University and Carnegie Mellon University*

In our paper that appeared in Vol. 1, No. 1 of *The Annals of Applied Statistics*, the second equation on page 24 should have an additional term $+ \mathrm{H}(q_d)$. The correct version of the equation is as follows:

$$\mathcal{L}(\boldsymbol{\mu}, \boldsymbol{\Sigma}, \boldsymbol{\beta}_{1:K}; \boldsymbol{w}_{1:D}) \geq \sum_{d=1}^{D} \mathrm{E}_{q_d}[\log p(\boldsymbol{\eta}_d, \boldsymbol{z}_d, \boldsymbol{w}_d | \boldsymbol{\mu}, \boldsymbol{\Sigma}, \boldsymbol{\beta}_{1:K})] + \mathrm{H}(q_d).$$

Computer Science Department
Princeton University
Princeton, New Jersey 08540
USA
E-mail: blei@cs.princeton.edu

Computer Science Department
Machine Learning Department
Carnegie Mellon University
Pittsburgh, Pennsylvania 15213
USA
E-mail: lafferty@cs.cmu.edu